\documentclass[twocolumn]{revtex4}

\begin{document}
 \title{The 2+1 charged black hole in topologically massive Electrodynamics}
\author{Tom\'as Andrade, M\'aximo Ba\~nados, Rafael Benguria}
\affiliation{Departamento de F\'{\i}sica,\\
 P. Universidad Cat\'olica de Chile, Casilla 306, Santiago 22,Chile,
\\
\\
 and}

 \author{Andr\'es Gomberoff}
 \affiliation{Centro de Estudios Cient\'{\i}ficos (CECS), Valdivia, Chile,
and\\
 Universidad Nacional Andr\'es Bello, Av. Rep\'ublica 237, Santiago, Chile.}

 \begin{abstract}

The 2+1 black hole coupled to a Maxwell field can be charged in
two different ways. On the one hand,  it can support a Coulomb
field whose potential grows logarithmically in the radial
coordinate. On the other, due to the existence of a
non-contractible cycle, it also supports a topological charge
whose value is given by the corresponding Abelian holonomy. Only
the Coulomb charge, however, is given by a constant flux integral
with an associated continuity equation. The topological charge
does not gravitate and is somehow decoupled from the black hole.
This situation changes abruptly if one turns on the Chern-Simons
term for the Maxwell field. First, the flux integral at infinity
becomes equal to the topological charge. Second, demanding
regularity of the black hole horizon, it is found that the Coulomb
charge (whose associated potential now decays by a power law) must
vanish identically. Hence, in 2+1 topologically massive
electrodynamics coupled to gravity, the black hole can only
support holonomies for the Maxwell field. This means that the
charged black hole, as the uncharged one, is constructed from the
vacuum by means of spacetime identifications.

\end{abstract}

\maketitle

The consistency of general relativity with quantum mechanics is
still one the most important problems of theoretical physics.
Since the problem was first formulated several decades ago, it
became clear that black holes would play a key role as a tool to
explore quantum gravity.  Two striking properties of these
particular solutions of Einstein equations --which have driven
most of the work on black hole quantum mechanics for many years--
are Hawking radiation and the so called ``no hair" theorems.

The non hair theorems (see \cite{hair} for a review) imply that
black holes are described only by their mass, angular momentum and
charge. In string theory, for example, charged objects have played
a key role in recent developments. D-branes and most extended
objects are supported by p-form fields carrying some conserved
charge. Black holes on these branes are also known, and in fact,
they had been the first examples for which an statistical
description of the Bekenstein-Hawking entropy is available
\cite{Strominger-}.

Black holes \cite{BTZ} also exists in the simpler setting of
three-dimensional gravity, and they share most of the properties
of the higher dimensional ones. Even though 2+1 gravity does not
contain gravitational waves, it is now clear that it encodes a
number of interesting properties \cite{CarlipBook}. To quote some
examples, the bold discovery by Brown and Henneaux \cite{BH} of a
centrally extended asymptotic conformal algebra, and its
Chern-Simons formulation \cite{Achucarro-}.

We shall be interested in this paper in the charged version of the
BTZ \cite{BTZ} black hole. In 2+1 dimensions, the action
describing the Maxwell field can be generalized to contain a
Chern-Simons (``topological mass") term \cite{Deser-}. We consider
then the action
\begin{equation}\label{action}
I = \int \sqrt{-g} \left( R + 2 - {\kappa \over 4}  F^{\mu\nu}
F_{\mu\nu} \right)  -{\alpha\over 2} \int \epsilon^{\mu\nu\rho}
A_\mu F_{\nu\rho}
\end{equation}
We stress that both, the Maxwell and Chern-Simons terms, are
quadratic in $A$ and gauge invariant.  There is thus no a priori
reason to exclude one in favor of the other. Furthermore, the
``massive" character of the Chern-Simons term implies that the
Brown-Henneaux \cite{BH} symmetry still applies in this theory, as
opposed to pure Maxwell theory.

In five and eleven dimensions, Chern-Simons terms arise as a
consequence of supersymmetry. Black hole solutions for the five
dimensional version of (\ref{action}) have been found recently in
\cite{Cvetic-}. As we shall see, in  three dimension there are
some surprises.

Particular black hole solutions to (\ref{action}) are known in
various cases. For $\kappa=\alpha=0$ one finds the uncharged BTZ
black hole \cite{BTZ}
\begin{eqnarray}
  ds^2 &=& -N^2(r) dt^2 + {dr^2 \over  N^2(r)}  + r^2\left( d\varphi + {J \over
2r^2} dt \right)^2  \label{BTZ0}
\end{eqnarray}
where
\begin{equation}
N^2(r) =  -M + r^2 + {J^2 \over 4r^2}.
\end{equation}
The asymptotic charges for this solution are $M$ and $J$. The
geometry is regular for all positive values of $r$, and there is a
regular event horizon provided $M > |J|$.

If $\kappa \neq 0 $ but still $\alpha=0$ one finds a more
complicated set of equations whose solution has been found in
\cite{MTZ} (this system was not fully solved in \cite{BTZ}).  The
black hole is now described by three asymptotic conserved charges
$M,J$ and the charge $C$. If $J=0$, the charged metric has the
form (\ref{BTZ0}) with $N^2(r) =  -M + r^2 + C^2 \log(r)$.

If $\kappa=0$ and $\alpha \neq 0$, the Abelian field decouples
from gravity. Globally, however, the gauge field can feel the
presence of the black hole by developing a non-zero holonomy
around the non-contractible loop. The general solution on this
case is the vacuum black hole (\ref{BTZ0}) supplemented with the
constant value for the gauge field,
\begin{equation}\label{Q}
A_{\varphi} = {Q \over 2\pi}, \ \ \ \  \Rightarrow \ \ \ \  \oint
A = Q.
\end{equation}
The ``topological charge" $Q$  cannot be eliminated by a
well-defined gauge transformation because $\varphi$ is compact
\footnote{The holonomy (\ref{Q}) is already present in the pure
Maxwell theory with $\kappa\neq 0$ and $\alpha=0$. In this case,
however, $Q$ is not related to the electric charge. See
\cite{Bowick-} for a related discussion in $d=4$ with a 2-form.}.
Furthermore, for $\kappa= 0$ and $\alpha\neq 0$, $Q$ is equal to
the electric charge of the system, defined as the conserved
quantity associated to non-trivial asymptotic gauge
transformations. It can also be represented as the zero mode of
the infinite-dimensional asymptotic $U(1)$ Kac-Moody algebra. We
shall prove in this paper that this configuration is stable
against the incorporation of the Maxwell term, that is turning on
the coupling $\kappa$.

Consider then the case $\kappa\neq 0$ and $\alpha\neq 0$. In what
follows we set $\kappa=1$. Selfdual \cite{Jackiw-}, particle-like,
solutions to this system were first discussed in \cite{Clement}
and further analyzed in \cite{Fernando-}. Our interest will be
focus on black holes solutions.

We start by writting the equations of motion. For our purposes it
will be convenient to use the following  parametrization for the
black hole ansatz,
\begin{eqnarray}
ds^2 &=& {dr^2 \over h^2 - p\, q} + p\, dt^2 + 2h\, dt d\varphi +
q\, d\varphi^2\label{ds1}
\end{eqnarray}
where $p,q,h$ are  functions of $r$ only. For the gauge field, we
write
\begin{eqnarray}
A_t(r) &=& \Phi(r), \label{At} \\
A_\varphi(r) &=& {Q \over 2\pi} + \chi(r)\label{Aphi}.
\end{eqnarray}
We have isolated the constant part of $A_\varphi$ (holonomy), and
assume that both $\Phi(r)$ and $\chi(r)$ vanish at infinity (this
is true only for $\alpha\neq 0)$.

Inserting (\ref{ds1},\ref{At},\ref{Aphi}) into Einstein equations
coupled to the Maxwell field and taking appropriated combinations
 we find the remarkable simple set of
equations (prime indicates radial derivative)
\begin{eqnarray}
  h'' &=& -\Phi'\, \chi' \label{1} \\
  p'' &=& -\Phi'^{\, 2} \label{2}\\
  q'' &=& -\chi'^{\, 2} \label{3}.
\end{eqnarray}
The equations also imply $(h^2-pq)''=h'^2-p'q' + 4$ (recall that
we are using $\Lambda=-1$). It is direct to see, however, that
(\ref{1}-\ref{3}) imply $(h^2-pq)'''= (h'^2-p'q')'$ and hence we
can omit this extra equation provided we fix the integration
constant as $4$.

The Maxwell-Chern-Simons equations become
\begin{eqnarray}
 h\Phi' - p \chi'  &=& 2\alpha \Phi, \label{4}\\
 q \Phi'-h \chi' &=& 2\alpha \chi . \label{5}
\end{eqnarray}
These equations are first integrals of the original ones. The two
integrations constants are the holonomy, $Q/2\pi$, represented by
the constant part of $A_\varphi$, and a constant added to $A_t$
which is trivial and can be gauged away.

The reduced set of equations (\ref{1}-\ref{5}) admits a Lagrangian
representation \cite{Clement},
\begin{equation}\label{L}
L = {1 \over 4} \mbox{Tr} \left( \Omega'^2 \right) + \alpha\bar
A\, A' - 2\alpha^2\, \bar A \Omega^{-1} A,
\end{equation}
where the functions $h,p,q$ and $\Phi,\chi$ are collected in the
$SL(2,\Re)$ matrix $\Omega$ and ``spinor" $A$,
\begin{equation}\label{Omega}
\Omega = \left(\begin{array}{cc}
  h & -p \\
  q & -h \\
\end{array}\right), \ \ \ \ \ A = \left(\begin{array}{c}
  \Phi \\
  \chi \\
\end{array}\right).
\end{equation}
The ``Dirac conjugate" is defined as $\bar A = A^t i =(\chi,-\Phi)
$ where $i$ is the real antisymmetric matrix in $2$ dimensions.
Eqns. (\ref{1}-\ref{3}) and (\ref{4}-\ref{5}) take the form
$\Omega'' = - A'\otimes \bar A'$ and $\Omega A ' = 2\alpha A$,
respectively, and can be derived by extremizing (\ref{L}).

The Lagrangian (\ref{L}) is invariant under $SL(2,\Re)$ rotations
(this symmetry is implicit in the ansatz (\ref{ds1}))
\begin{equation}\label{U}
\Omega \rightarrow U^{-1} \Omega U, \ \ \ \ \  A \rightarrow
U^{-1} A, \ \ \ \ \  \bar A \rightarrow \bar A U
\end{equation}
where $U$ is a {\it constant} matrix with $\det(U)=1$. The
corresponding Noether charge, satisfying $K'=0$, is given by,
\begin{equation}\label{K}
K ={1 \over 2}[\Omega,\Omega'] + \alpha A\otimes \bar A .
\end{equation}
The Lagrangian (\ref{L}) is also invariant under constant
translations in the radial coordinate $r\rightarrow r+a$ and the
associated Noether charge is the ``energy",
\begin{equation}
{\cal E} ={1 \over 4} \mbox{Tr} \left(\Omega'^2 \right) +
2\alpha^2\,\bar A \Omega^{-1}A.
\end{equation}
This integral is however not an arbitrary constant but it is fixed
by Einstein equations. In fact, it is direct to prove that $2{\cal
E} = (h^2-pq)'' - (h'^2-p'q')$. Hence, according to the discussion
after Eq. (\ref{3}), we must fix ${\cal E} = 2$.

The vacuum solution (BTZ black hole) in this representation is
given by $\Phi=0=\chi$ and,
\begin{eqnarray}
  h(r) &=& J, \nonumber\\
  p(r) &=& 4(M-r), \label{BTZ1}\\
  q(r) &=& r.  \nonumber
\end{eqnarray}
In fact, replacing (\ref{BTZ1}) in (\ref{ds1}) and making the
radial redefinition $r \rightarrow r^2$ one obtains (\ref{BTZ0}).

We would like to add charge to (\ref{BTZ1}).  We first study the
asymptotic structure of the charged solution. Setting (\ref{BTZ1})
as background, we develop a series expansion in powers of $1/r$
for the solution. This can be done to any desired order. We
display here the first few terms,
\begin{eqnarray}
  h &=& J + {C^2 \alpha \over 4(2\alpha+1)}{1 \over r^{2\alpha}}\left(1 +
{M\alpha \over r} + \cdots  \right),\nonumber\\
  p &=&  4(M-r) - {C^2 \alpha \over 2(2\alpha+1)}{1 \over r^{2\alpha}}\left(1 +
 {2M\alpha-J \over 2\alpha+1} \, {\alpha \over r} +\cdots \right), \nonumber\\
  q &=&  r - {C^2 \alpha \over 8(2\alpha+1)}{1 \over r^{2\alpha}}\left( 1 +
{2M\alpha + 2M +J \over 2\alpha+1}\, {\alpha \over r} +\cdots
\right),
\nonumber\\
  \Phi &=& {C \over r^{\alpha}} \left( 1 + {2M\alpha-J \over 2(2\alpha+1)}
{\alpha \over r}  + \cdots            \right), \nonumber\\
  \chi &=&  -{C \over 2r^{\alpha}} \left( 1 + {(J + 2M\alpha + 2M) \over
2(2\alpha+1)} {\alpha \over r}  + \cdots \right).\nonumber
\end{eqnarray}
Here, $C$ is an integration constant which will be called
``Coulomb Charge".

There exists another solution whose gauge field diverges as
$r^{\alpha}$, and hence we discard it. We assume $\alpha>0$. Note
that the $C\,$log$(r)$ structure arising in pure Maxwell theory
has been replaced by $C/r^{\alpha}$. This is a consequence of the
``massive" character of the Chern-Simons term.

The asymptotic solution is thus characterized by four parameters,
the mass $M$, angular momentum $J$, topological charge $Q$, and
Coulomb charge $C$.  Let us prove that the associated electric
charge is equal to $Q$.

In the presence of the Chern-Simons term, the definition of
electric charge has some subtleties. This problem was first
studied in \cite{HT}. See \cite{Townsend-} for recent discussions.
For the case at hand, the electric charge of the system is
\cite{HT}
\begin{equation}
\int_{\gamma} \left( {}^*\! F + 2\alpha A \right) \label{charge}
\end{equation}
where the integral encloses the origin $r=0$ and it is assumed to
be outside sources. As usual, the value of the integral does not
depend locally on $\gamma$. We choose the curve to be a circle of
a large radius at some constant time. From the above ansatz we
find ${}^*\!F_\phi = -q\Phi'+h\chi'$ and hence the charge becomes,
\begin{eqnarray}
\int d\phi \left[ -q\Phi'+h\chi' + 2\alpha \left(\frac{Q}{2\pi} +
\chi \right) \right]
 = 2\alpha Q, \label{Q1}
\end{eqnarray}
where we have used the equation of motion (\ref{5}). We thus find
that the electric charge of the system is proportional to the
topological charge $Q$. The Coulomb charge $C$ is not related to
an asymptotic symmetry, and it would appear as ``hair" in the
corresponding black hole solution.

We now turn into the problem of finding black hole solutions to
(\ref{1}-\ref{5}), satisfying the asymptotic conditions found
above.  To analyze the horizon geometry we rewrite the ansatz
(\ref{ds1}) in its ``ADM" form
\begin{equation}\label{ADM}
ds^2   = - {h^2 - pq \over q} dt^2 + {dr^2 \over h^2-pq} + q
\left( d\varphi + {h \over q} dt\right)^2 .
\end{equation}

From here we see that the structure of horizons is controlled by
the function,
\begin{equation}\label{fdef}
f(r) \equiv h^2 - pq
\end{equation}
A point $r=r_+$ satisfying the following three conditions: (i)
$f_+=0$, (ii) the horizon area $q_+$ is not zero and, (iii) the
angular velocity of the horizon ${h_+\over q_+}$ is not singular,
defines a regular event horizon. Here and below, a subscript $+$
means the corresponding function evaluated at the horizon.

Although the system (\ref{1}-\ref{5}) does not look too
complicated, its solution for all values of $r$ has escaped us.
(See \cite{Clement} for a particular solution describing
``particles"). Nevertheless, without knowledge of the exact
solution, we shall be able to prove that a regular horizon can
exists if and only if the Coulomb charge vanishes identically,
\begin{equation}
C=0.
\end{equation}
To this end, we first prove that the function $q$ must be positive
for all $r>r_+$. Since $q_+$ is the horizon area, we have $q_+>0$.
{}From (\ref{3}) we observe that $q^{\prime\prime}$ is negative
for all $r$. If $q_+'<0$ then $q(r)$ would eventually become
negative, in contradiction with its asymptotic value given above.
Thus, $q_+'$ must be positive and, as a consequence, $q$ is
positive for all $r>r_+$.

In a similar way, we now prove that $p$ must be negative for all
$r>r_+$. First, we note that one can always choose the angular
velocity at the horizon to be zero, that is, $h_+=0$. This is
achieved by the transformation $\varphi \rightarrow \varphi +
w\,t$ which, acting on (\ref{ADM}), has the effect of shifting
$h/q \rightarrow h/q +w$ \footnote{Since the angular velocity at
the horizon is now zero, one must expect the angular velocity at
infinity to be different from zero. Its value can be found by
computing the matrix $K$, which does not depend on $r$, (see Eqn.
(\ref{K})) at infinity and at the horizon. The only relevant point
for us is that the transformation $\varphi \rightarrow \varphi +
wt$ leaves $q$ and $A_\varphi$ invariant.}. From now on, we work
on this frame. Next, we note that since $h_+^2-p_+q_+=0$ and $q_+$
is finite, it follows that $p_+=0$. This implies that
\begin{equation}
f_+'= 2h_+h_+'-p_+'q_+-p_+q_+'= -p_+'q_+. \label{sign}
\end{equation}
The function $f$ is positive for $r>r_+$ and vanishes at $r=r_+$.
This implies that $f_+'\geq 0$ and since $q_+>0$ we find that
$p'_+$ must be negative or zero. From Eq. (\ref{2}) we see that
$p^{\prime\prime}$ is negative for all $r$. We conclude that $p'$
must be negative all the way from $r_+$ to infinity. Since $p$
vanishes at the horizon, this means that $p$ must be negative for
all $r>r_+$.

We now analyze the behavior of the gauge field. First we note that
$\Phi_+=0$. This follows directly from Eqn. (\ref{4}), with
$\alpha\neq 0$. On the other hand, from the asymptotic series
displayed above, we see that both functions $\Phi$ and $\chi$
vanish as $r\rightarrow\infty$. Then, the function $\sigma \equiv
\Phi\chi$ vanishes both at the horizon and at infinity.  (We
assume here that $\chi_+$ is finite, which is required for
regularity of the horizon  \footnote{In a gauge theory one should
only impose regularity on the field strengths,  not the gauge
potentials. However, in this case the conserved quantity $K$ again
imply that $\chi_+$ must be regular at the horizon for the
geometry to be well defined. }.)  Since $\sigma$ must be
continuous we find that either $\sigma$ is zero everywhere or has
an extremum at some value $r=r_0$. If the latter is true, then the
derivative of $\sigma$ at each side of $r_0$ has different sign,
and therefore there must be a region where $\sigma'$ is negative.
But this is contradiction with the equation of motion. In fact,
multiplying Eq. (15) by $\chi'$ and Eq (16) by $\Phi'$, and adding
them together, one obtains
\begin{equation}
q(\Phi')^2 - p (\chi')^2 = 2\alpha\sigma' \ .
\end{equation}
The left hand side of this expression is non-negative, because
$p<0$ and $q>0$ for all $r>r_+$. Therefore $\sigma'$ cannot take
negative values. We conclude that $\sigma$ must be identically
zero everywhere and thus either $\Phi$ or $\chi$ must be zero. By
inspection of the asymptotic solution this imply that $C$ must
vanish. The solution thus reduces to the BTZ black hole ``charged"
with the holonomy $Q$, as anticipated.

To summarize, we have shown in this paper that BTZ black holes
coupled to Maxwell-Chern-Simons electrodynamics can support
holonomies for the gauge field, but not a local electromagnetic
field. A star could support this field and we have found its
explicit asymptotic form. Upon gravitational collapse, however,
the field must be expelled before a regular horizon could be
formed. We hope to come back to this problem in the future. A more
detailed analysis of the results presented here will be presented
elsewhere.

We would like to thank E. Ay\'on-Beato, G. Dunne, M. Henneaux and
A. Faraggi, for useful insights and comments during the curse of
this work. MB was partially supported by Fondecyt grants \#
1020832 and \# 7020832. RB was partially supported by Fondecyt
grant \# 1020844. AG was partially supported by FONDECYT grants \#
1010449 and \# 1010446. Institutional grants to CECS of the
Millennium Science Initiative, Fundaci\'on Andes, and the generous
support by Empresas CMPC are also gratefully acknowledged.

 \end{document}